\documentstyle[aps,manuscript]{revtex}

\begin{document}

\title{Probing equilibration with respect to isospin
degree of freedom in intermediate energy heavy ion collisions
\footnote{Supported by National
Natural Science Foundation of China under Grant No. 19975073, 
and Science Foundation of Nuclear Industry and the Major
State Basic Research Development Program under the contract No. G20000774 }}

\author {Qingfeng Li$^{1)}$, Zhuxia Li$^{1,2,3)}$}
\address{ 1) China Institute of Atomic Energy, P. O. Box 275 (18), 
Beijing 102413, P. R. China\\
2) Center of Theoretical
Nuclear Physics, National Laboratory of Lanzhou Heavy Ion Accelerator, 
 Lanzhou 730000, P. R. China\\
3) Institute of Theoretical Physics, 
Academia Sinica, P. O. Box 2735, Beijing 100080, P. R. China }
\maketitle

%\date{\today}

\begin{abstract}
 We have studied equilibration with respect to isospin degree of freedom
 in four 96 mass systems $^{96}Ru+^{96}Ru$, $^{96}Ru+^{96}Zr$, 
 $^{96}Zr+^{96}Ru$, $^{96}Zr+^{96}Zr$ at 100 AMeV and 400 AMeV with isospin dependent QMD. 
 We propose that the neutron-proton differential rapidity distribution 
is a sensitive probe to the degree of equilibration with respect to the isospin 
degree of freedom. By analyzing the average $N/Z$ ratio of 
emitted nucleons, light charged particles (LCP) and intermediate mass fragments (IMF), 
it is found that there exist memory effect in multifragmentation process. 
The average $N/Z$ ratio of IMF reduces largely as beam energy increases from 100 AMeV to 400 AMeV 
which may result from the change of the behavior of the isotope distribution of IMF charges. 
The isotope distribution of IMF charges does also show certain memory effect at 100 AMeV case 
but not at 400 AMeV case. 
\end{abstract}
%\pacs{PACS: 25.70.Pq, 24.10.-i.}
PACS numbers: 25.70.Pq, 24.10.-i\newline

The study of whether the equilibrium is reached or not is a prerequisite 
for the extraction of valid information about the thermodynamical properties 
of the excited system produced in the reaction. This problem has been studied 
theoretically 
and experimentally for many years. But still there is many new problems which 
need to further study. 
Especial interest is about the nature of the multifragmentation that is if  
the multifragmentation is a statistical emission process or the dynamical one
\cite{Ai00,Bon95,Gr93,Ai91,Go97}. To clarify this problem, the FOPI
collaboration recently performed a so-called 'mixing experiment' using four 
mass $96+96$ systems $Ru+Ru$, $Zr+Zr$, $Ru+Zr$, $Zr+Ru$ at 400 AMeV\cite{FOPI00,Re00}. 
To quantify conveniently the 'degree of mixing', they defined a normalized 
proton counting by the value of $Zr+Zr$ and $Ru+Ru$,
\begin{equation}
R_{Z}=\frac{2*Z-Z^{Zr}-Z^{Ru}}{Z^{Zr}-Z^{Ru}}.
\label{eq1}
\end{equation}
They first measured the proton counting number for $Ru+Ru$ and $Zr+Zr$ then they
measured $R_{Z}$ for asymmetric reaction $Zr+Ru$. 
The results of $R_{Z}$  for reaction $Zr+Ru$ showed that the protons were
not emitted
from an equilibrium source and the reaction was half transparent\cite{FOPI00}. 
These experimental results told us that the equilibrium was not eventually
reached in the reaction. However, 
this beautiful experiment study has only shown that at beam energy 400 AMeV
the protons are emitted by a non-equilibrium source but it still can not answer
if multifragmentation is a statistical emission process or dynamical emission
one at lower energy.
 
The aim of this work is to test the non-equilibrium effect 
by means of isospin degree of freedom relevant probes stimulated by the 'mixing experiments'
performed by FOPI collaboration. We will first introduce our model briefly then
we study the normalized proton counting  $R_{Z}$ and other probes such as the proton
rapidity distribution, neutron-proton differential rapidity distribution, and the 
isospin distribution of emitted nucleons, light charged particles
and intermediate mass fragments in the same collision systems at 400 AMeV as well as
100 AMeV. And finally a short conclusion will be given.

The Isospin Dependent Quantum
Molecular Dynamics (QMD) model\cite{Ai91,Har89} is used in the calculations. 
The following modifications in QMD model are introduced: Firstly, the isospin 
dependent part of the nuclear potential is taken into account
in addition to the
Coulomb interaction. The symmetry potential energy per nucleon is taken as the following form, 
\medskip 
\begin{equation}
V_{sym}(\rho, \delta )=\frac{C_S}{2} (\frac{\rho}{\rho_{0}}) \delta^2,  \label{eq2}
\end{equation}
where
\begin{equation}
\delta =\frac{\rho_n-\rho_p}{\rho_n+\rho_p},  \label{eq3}
\end{equation}
and $C_S$ is the strength of symmetry potential energy. In this work it is taken to be 35 MeV and the 
corresponding symmetry energy is about 29 MeV. 
Secondly, the isospin dependent binary elastic scattering cross section 
is used. It is well known that
up to hundreds MeV the free elastic proton-neutron cross section
is about 2-3 times larger than that of proton-proton (neutron-neutron)'s. 
Finally, in
the treatment of the Pauli blocking, we firstly distinguish protons and
neutrons, and then we use the following two criteria:
\begin{equation}
\frac{4\pi}{3} r_{ij}^{3}\cdot \frac{4\pi}{3}p_{ij}^{3}\geq
\frac{h^{3}}{4},  \label{eq4}
\end{equation}
and
\begin{equation}
P_{block}=1-(1-f_{i})(1-f_{j}),  \label{eq5}
\end{equation}
where $f_{i}$ is the distribution function in phase space for particle $i$
and reads as
\begin{equation}
f_i(\stackrel{\rightarrow }{r},\stackrel{\rightarrow }{p},t)=\frac {1}{\pi
\hbar^{3}}\exp (-(\stackrel{\rightarrow }{r}-\stackrel{\rightarrow }{r_i}
(t))^{2}/2L^{2})\exp (-(\stackrel{\rightarrow }{p}-\stackrel{\rightarrow }
{p_i} (t))^{2}2L^{2}/\hbar^{2}).  \label{eq6}
\end{equation}
Where $L$ is a parameter which represents the spatial spread of wave packet, 
$\stackrel{\rightarrow }{r_i}(t)$ and $\stackrel{\rightarrow }{p_i}(t)$
denote the center of the wave packet in coordinate and momentum space
respectively. The first condition gives the criterion for the uncertainty relation of 
the centroids of Gaussion
wave packets of two particles. The second one is the probability of the 
Pauli blocking effect for the scattering of two particles, which is
especially useful for collisions of heavy nuclei. 
The soft EOS ($K=200MeV$) is used in the calculations, the
corresponding main parameters are listed in Table 1. The secondary deexcitation on primary 
hot fragments is not taken into account in the present calculations. It should not 
change the general conclusion of this work.

We firstly investigate the proton counting for the mixing reactions of 
four mass $96+96$ systems $Ru+Ru$, $Zr+Zr$, $Ru+Zr$, $Zr+Ru$ as the same as in
the experimental
study in ref.\ \cite{FOPI00}. According to the definition of $R_{Z}$, 
$R_{Z}=1$ for $Zr+Zr$, $R_{Z}=-1$ for $Ru+Ru$. 
For asymmetric reaction $Ru+Zr$, $Zr+Ru$, it may be more convenient to express  $R_{Z}$  as  
$R_{z}=2R_{mix}-1$, for $Zr+Ru$ and $R_{z}=1-2R_{mix}$ for $Ru+Zr$, which can be derived 
from definition (1). Here $R_{mix}$ is the percentage of 
the number of protons emitted from projectile. 
$R_{mix}$ is proportional to the degree of mixing of projectile and target.  
It is obvious that if projectile and target is completely mixed then $R_{mix}$ equals to 0.5 
at any rapidity and if the reaction is full transparent $R_{mix}$ should equal to 1 at 
projectile rapidity and 0 at target rapidity, respectively. 
Fig. 1 shows $R_Z$ as function of rapidity at beam energy 100 AMeV impact 
parameter $b=0fm$ 
and 400 AMeV $b=0fm$ and $b=5fm$. The experimental data (at 400 AMeV) is also
given in the figure. From this figure, one
can easily find that the absolute $R_{Z}$ value goes from zero to about 0.5 for reactions 
$Zr+Ru$ and $Ru+Zr$ at energy 100 AMeV and 400 AMeV $b=0fm$, and about 0.75 for 
the same reactions at beam energy 400 AMeV, $b=5fm$. Our calculation is in
reasonable agreement with experiments data and consequently, 
the same conclusion concerning the non-equilibrium effect can be
drawn for 400 AMeV case. The results for $b=0fm$ and $b=5fm$ show 
the non-equilibrium effect strongly depends on the impact parameter. 
However, the results of $R_{Z}$ for 400 AMeV and 100 AMeV at $b=0fm$ are
undistinguishable and they lead to the same conclusion that the protons are produced in
a non-equilibrium source at both 400 AMeV and 100 AMeV. 
It seems to us that $R_{Z}$ is not very sensitive to  
the energy dependence of the mixing of projectile and target at the energy range studied 
in this work. We also find that $R_{Z}$ is also not sensitive to the symmetry potential, which
will be discussed in our another work. 
In the following we make further investigation in order to find other possible probes which 
may provide more clear information for the energy dependence of the degree of equilibrium.

\[
\fbox{Fig. 1 } 
\]

In Fig. 2 a)-b) we show the rapidity distribution of emitted protons 
at beam energy 100 AMeV and 400 AMeV. 
From Fig. 2 a) and b) we can find that the reaction 
$^{96}Ru+^{96}Ru$ emits more protons than that of $^{96}Zr+^{96}Zr$ 
because of 8 protons difference for two reaction systems. 
The proton rapidity distribution for
$^{96}Zr+^{96}Ru$ and $^{96}Ru+^{96}Zr$ is between those of $Ru+Ru$ 
and $Zr+Zr$. Differing from the symmetric reaction $Ru+Ru$ and $Zr+Zr$, 
the rapidity distribution of emitted protons for $Ru+Zr$ and $Zr+Ru$ 
is asymmetric and the peaks deviate from $Y=0$. 
It again means that the protons are emitted from
a non-equilibrium source. But again we find it is difficult to give clear 
energy dependence of the degree of equilibrium reached.  
As we know that comparing with the
most stable isotopes $^{102}Ru$ and $^{90}Zr$, $^{96}Ru$ is of 6
neutron deficiency and $^{96}Zr$ is of 6 neutron excess. The ratio between
proton number and neutron number for $^{96}Ru$ and $^{96}Zr$ is 0.85 and 0.71, 
respectively. It would be more desirable to study the rapidity distribution of
the isovector density of emitting nucleons for isospin asymmetric nuclear systems. 
Therefore we introduce the neutron-proton differential rapidity distribution. 
Fig. 3 a), b) and c) show the
neutron-proton differential rapidity distribution
for $^{96}Ru+^{96}Ru$, $^{96}Zr+^{96}Zr$, $^{96}Zr+^{96}Ru$, $^{96}Ru+^{96}Zr$
at a) 100 AMeV, $b=0fm$, b) 400 AMeV, $b=0fm$  and c) 400 AMeV, $b=5fm$.  
First, for all three cases a), b) and c), the centroids of neutron-proton differential 
rapidity distribution for
$^{96}Ru+^{96}Zr$, $^{96}Zr+^{96}Ru$ are located at the side of Zr 
(as target or projectile) and strongly deviate from $Y=0$. 
The centroid of distribution should be at $Y=0$ if a system is in equilibrium. 
The deviation of the centroid of neutron-proton differential rapidity 
distribution from $Y=0$ means there is non-equilibrium effect. The larger the deviation
from $Y=0$ is the stronger the non-equilibrium effect is. 
The deviation of the centroid of neutron-proton differential rapidity 
distribution from $Y=0$ for $b=5fm$ case is much larger than that for $b=0fm$ case. 
This is of course quite understandable.  
Further more, one can find that the neutron-proton differential rapidity
distribution of symmetric reactions
$^{96}Ru+^{96}Ru$, $^{96}Zr+^{96}Zr$ at 100 AMeV deviates from the Gaussion
shape more strongly than that at 400 AMeV. 
It implies that there exist obvious non-equilibrium effect in the emitting nucleon process. 
Therefore we can conclude that the neutron-proton differential rapidity 
distribution is a sensitive probe to explore the energy dependence of the degree of equilibrium  
for an isospin asymmetric system. We may generalize the neutron-proton differential
rapidity distribution by introducing t-$^{3}He$ differential rapidity
distribution to probe equilibration in isospin asymmetric colliding systems.  

\[
\fbox{Fig. 2 a), b)} 
\]

\[
\fbox{Fig. 3 a), b), c)} 
\]
However, emitted single nucleons can only characterize limited part of the system,  
therefore we further study the isospin distribution in LCP and IMF in addition to nucleons. 
In Fig. 4 (I) and (II), we show the average $N/Z$ ratios in emission of nucleons, LCP and  
IMF at projectile (a)), central (b)) and target (c)) 
rapidity region in four systems at 400 AMeV and 100 AMeV, $b=0fm$, respectively. 
The projectile rapidity region 
is defined by $1.5\geq Y\geq 0.5$, the target rapidity region 
$-0.5\geq Y\geq -1.5$
and the central rapidity region $0.5 \geq Y\geq -0.5$.  
The figures firstly tell us a basic feature that 
the difference between the average $N/Z$ ratios of emitted nucleons of 4 colliding systems 
with different isospin asymmetry is much larger than that between the average ratios of 
 LCP and IMF of 4 systems at three rapidity regions, i.e., 
the more neutron (proton) rich systems emit more neutrons (protons) 
while the average $N/Z$ ratios of LCP and IMF for these four
systems are relatively close. This behavior is stronger at 100 AMeV case. 
The experimental measurements at tens AMeV energy region
\cite{Ve00} found that the more asymmetric the
system is the stronger the system will be breaking up into still more neutron rich 
(deficient) light fragments while the $N/Z$ ratio of heavier fragments
remains relatively insensitive.  Our calculation results show similar tendency, only because 
of the energy difference, here the $N/Z$ ratios of emitted nucleons, LCP and IMF are compared 
instead of comparing the $N/Z$ ratios for
LCP and IMF in \cite{Ve00} where the energy was relatively low. 
The second feature is that the average $N/Z$ ratio of emitted nucleons generally is the
largest, and then, that of LCP and the $N/Z$ ratio of IMF is the smallest in all
rapidity region, which implies that the single nucleons are more neutron rich and  LCP and IMF 
are more isospin symmetric. 

It is more meaningful to investigate whether the $N/Z$ ratio of IMF for mixing reaction
converges or not as far as the degree of equilibrium is concerned because IMF 
produce at late stage of reaction\cite{Bo94}. 
When we attend to the $N/Z$ ratios at target and  projectile rapidity region, we
find that not only the $N/Z$ ratios of emitted nucleons for two mixing reactions 
$^{96}Zr+^{96}Ru$ and $^{96}Ru+^{96}Zr$ but also those of LCP and IMF do not merge each other 
but they are more close to $Zr+Zr$ or $Ru+Ru$ at respective rapidity region. 
It means that not only the nucleons but also IMF are not emitted 
from an completely equilibrium source. 
Of course, the difference of the average $N/Z$ ratio of IMF for reactions $Zr+Ru$ and $Ru+Zr$ is weaker 
than emitted nucleons, which is understandable because IMF produce at later stage. 
One can further find that the difference of the $N/Z$ of IMF for $Zr+Ru$ and $Ru+Zr$ at 100 AMeV 
is larger than that at 400 AMeV. It may also imply the energy dependence of the degree of equilibrium 
with respect to the isospin degree of freedom. The energy dependence of the degree of equilibrium
is because the two-body collisions become more violent as energy increases from 100 AMeV 
to 400 AMeV.  

By comparing Fig. 4 (I) and (II), one can find  that the $N/Z$ ratio
decreases as energy increases from 100 AMeV to 400 AMeV for all 4 systems. 
It would be interesting to study the reason of this behavior.  In Fig. 5 we show the 
yields of the isotopes of the most abundant IMF charges, a) Li, b) Be, c) B for $Zr+Zr$, $Zr+Ru$, 
$Ru+Zr$, $Ru+Ru$ 
at 100 AMeV and  $Zr+Zr$, $Ru+Ru$ at 400 AMeV, respectively. One can easily find that the yields of 
isotopes of Li, Be, B for 100 AMeV case are about several times larger (for non-neutron rich 
isotopes) to several tens times larger (for neutron rich isotopes) than those for 400 AMeV case. 
The curves for isotope distribution of Li, Be, B for 100 AMeV are more flat than those for
400 AMeV and furthermore the most abundant isotopes of Li, Be, B are always those of 
most stable ones for 100 AMeV case while are always those of the lightest isotopes for 400 AMeV 
case. Consequently, the average $N/Z$ of IMF is reduced as energy increases from 100 AMeV
to 400 AMeV. The another obvious difference between the isotope distribution of Li, Be, B for 
100 AMeV and 400 AMeV case is the dependence of the yields of the neutron rich (deficient) 
isotopes on the initial system. For 100 AMeV case, the relative yields of the neutron rich 
(deficient) isotopes depends on the $N/Z$ ratio of initial system. The initial system with larger 
$N/Z$ ratio produces more neutron rich isotopes and vice versa. 
We notice that for this case (100 AMeV) that the   
curves of the isotope distribution of Li, Be, B in mixing reactions $Zr+Ru$ and $Ru+Zr$ 
do not merge into one curve but they are close to those of respective reactions $Zr+Zr$ or $Ru+Ru$ with 
the same projectile. It means that there exist certain memory effect at 100 AMeV case. 
For 400 AMeV case, this memory effect appearing in the isotope distribution
of IMF charges disappears.       
     
\medskip 
\[
\fbox{Fig. 4} 
\]

\[
\fbox{Fig. 5} 
\]

In summary, we have studied the isospin relevant probes: normalized proton counting $R_{Z}$, 
the proton rapidity
distribution, the neutron-proton differential rapidity distribution as well as 
the $N/Z$ ratio of single nucleons, LCP, IMF at central, projectile and target rapidity region
for four 96 mass systems
$^{96}Ru+^{96}Ru$, $^{96}Ru+^{96}Zr$, $^{96}Zr+^{96}Ru$, $^{96}Zr+^{96}Zr$
at 100 AMeV and 400 AMeV with isospin dependent QMD. 
All these probes concerning the single nucleon emission studied in this work show 
that the emitted nucleons
are not from an equilibrium source and there exits obvious non-equilibrium
effect. We propose that the neutron-proton differential rapidity distribution
is a sensitive probe to the energy dependence of the degree of equilibrium in single nucleon 
emission in intermediate energy heavy ion collisions.  
The average $N/Z$ ratios of IMF in mixing reactions 
$^{96}Ru+^{96}Zr$, $^{96}Zr+^{96}Ru$ with the same $N/Z$ and mass do not converge but
they are more close to $Zr+Zr$ or $Ru+Ru$ at respective rapidity region. The difference of $N/Z$ 
ratios of IMF between $^{96}Ru+^{96}Zr$, $^{96}Zr+^{96}Ru$ at 100 AMeV is larger than that 
at 400 AMeV which shows the energy dependence of the non-equilibrium effect with respect to 
isospin degree of freedom.   
Furthermore, we find the average $N/Z$ ratios of IMF at projectile and target rapidity region 
of IMF largely decreases as energy increases from 100 AMeV to 400 AMeV, which may result 
from the change of the behavior of the isotope 
distribution of IMF charge from 100 AMeV to 400 AMeV. The analyzing of isotope distribution 
of IMF charges at projectile rapidity region for four 96 mass systems shows existing of 
memory effect at 100 AMeV but not at 400 AMeV concerning isospin degree of freedom.      
 
\acknowledgments
The author Z. Li thanks the research group for hadron science of Advanced
Science Research Center of JAERI for the hospitality, The manuscript is
partly finished during her visit there.

\begin{table}
\caption{ Parameters used in calculations}

\begin{tabular}{|c|c|c|c|c|c|c|}
\hline
$\alpha (MeV)$ & $\beta (MeV)$ & $\gamma $ & $\rho _0 (fm^{-3})$ & $K (MeV)$ & 
$L (fm)$ & $C_{Yuk} (MeV)$ \\ \hline
$-356$ & $303$ & $7./6.$ & $0.168$ & $200$ & $2.1$ & $-5.5$ \\ \hline
\end{tabular}
\end{table}

%\newpage

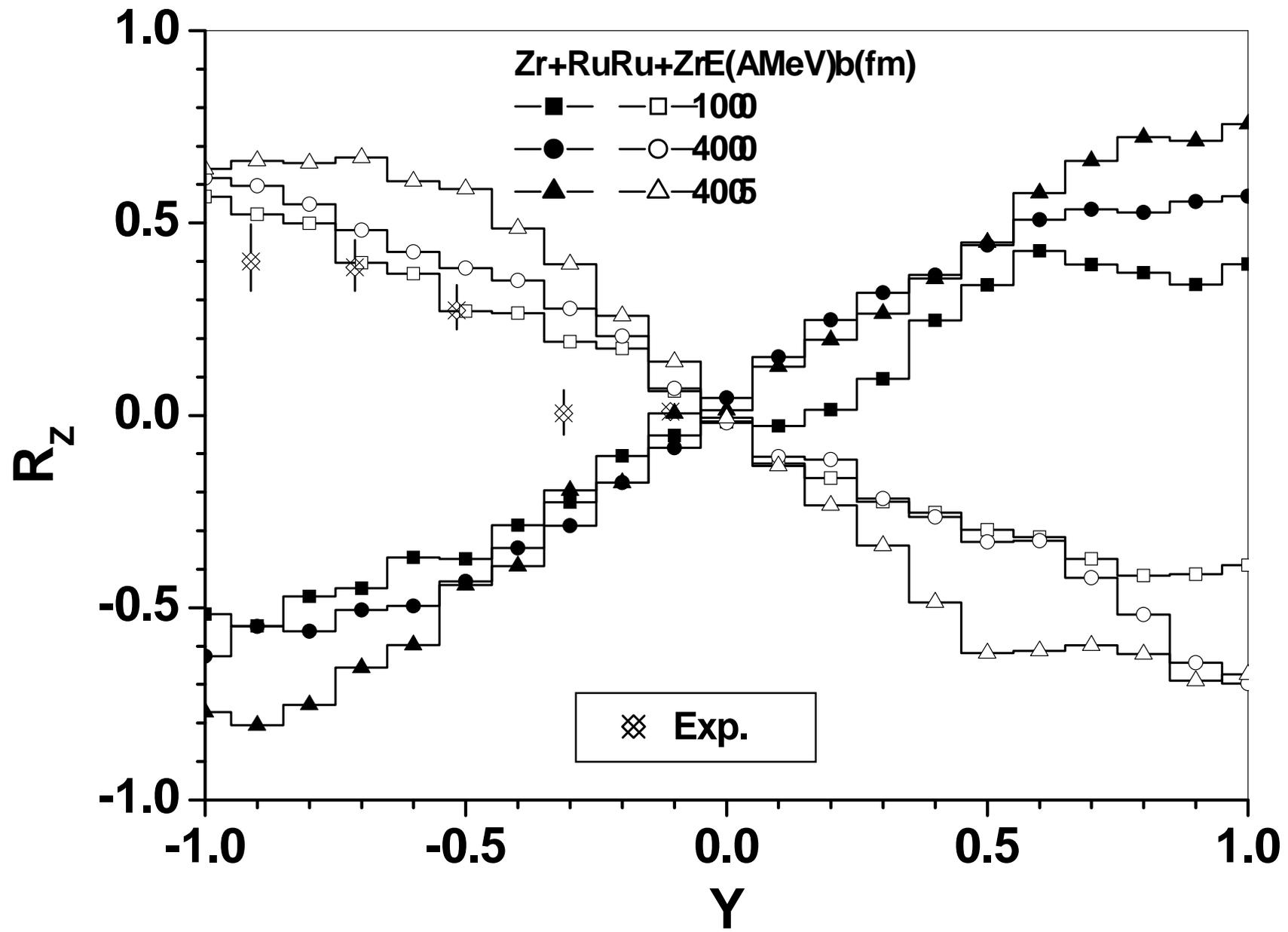
\begin{figure}
\caption{  The proton counting number R$_{z}$ as function of
            rapidity for
            $^{96}Ru+^{96}Ru$, $^{96}Ru+^{96}Zr$, $^{96}Zr+^{96}Ru$, 
            $^{96}Zr+^{96}Zr$
            at $E=100AMeV$ $b=0fm$ (a)), $E=400AMeV$ $b=0fm$ (b)) and $b=5fm$ (c)), respectively. 
            The experimental data for 400 AMeV
            are also given in the figure.
        }
\label{fig1}
\end{figure}

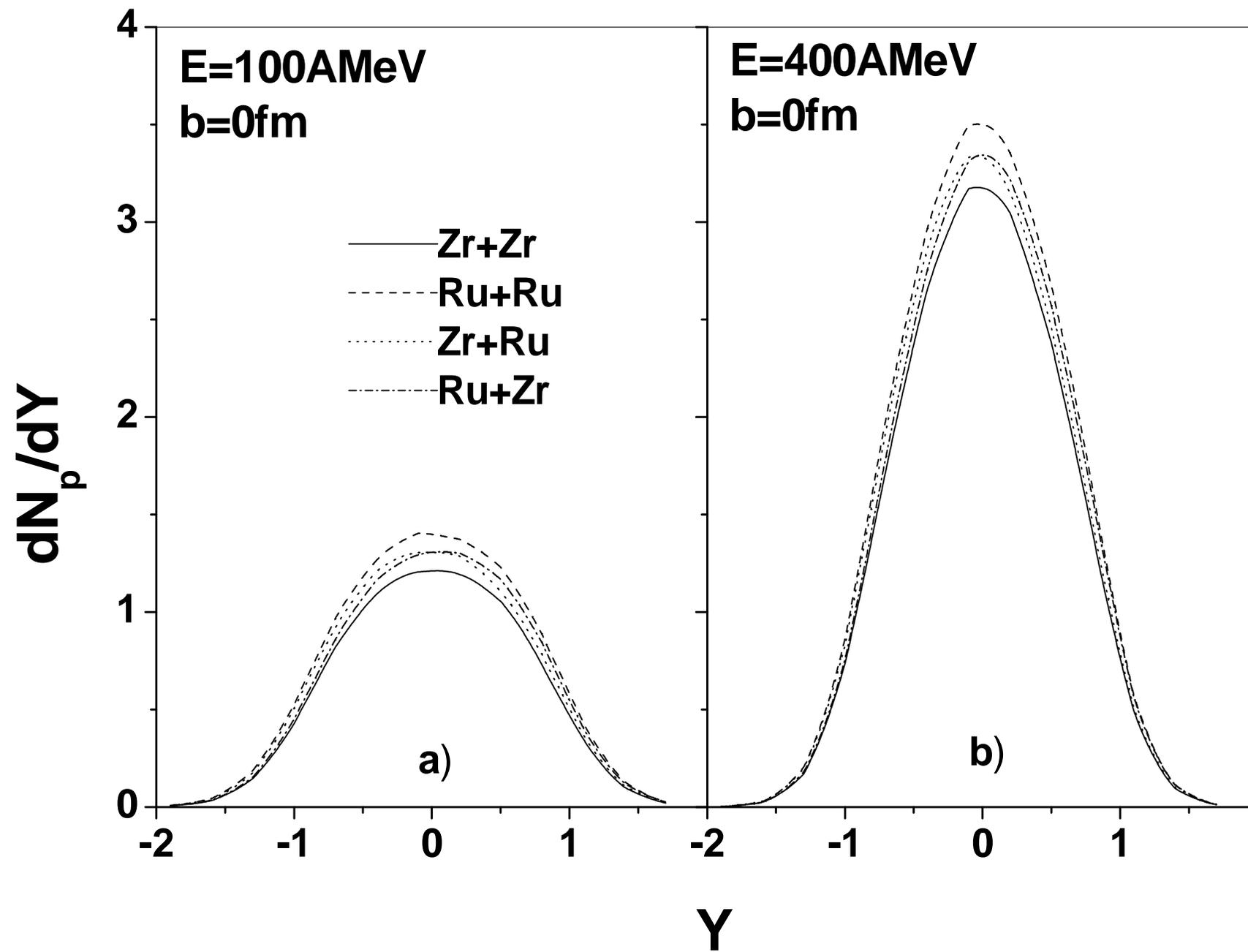
\begin{figure}
\caption{ The rapidity distribution of emitted protons for the same reaction
            as Fig. 1
               a) at 100 AMeV, $b=0fm$ and b) at 400 AMeV, $b=0fm$.
        }
\label{fig2}
\end{figure}  

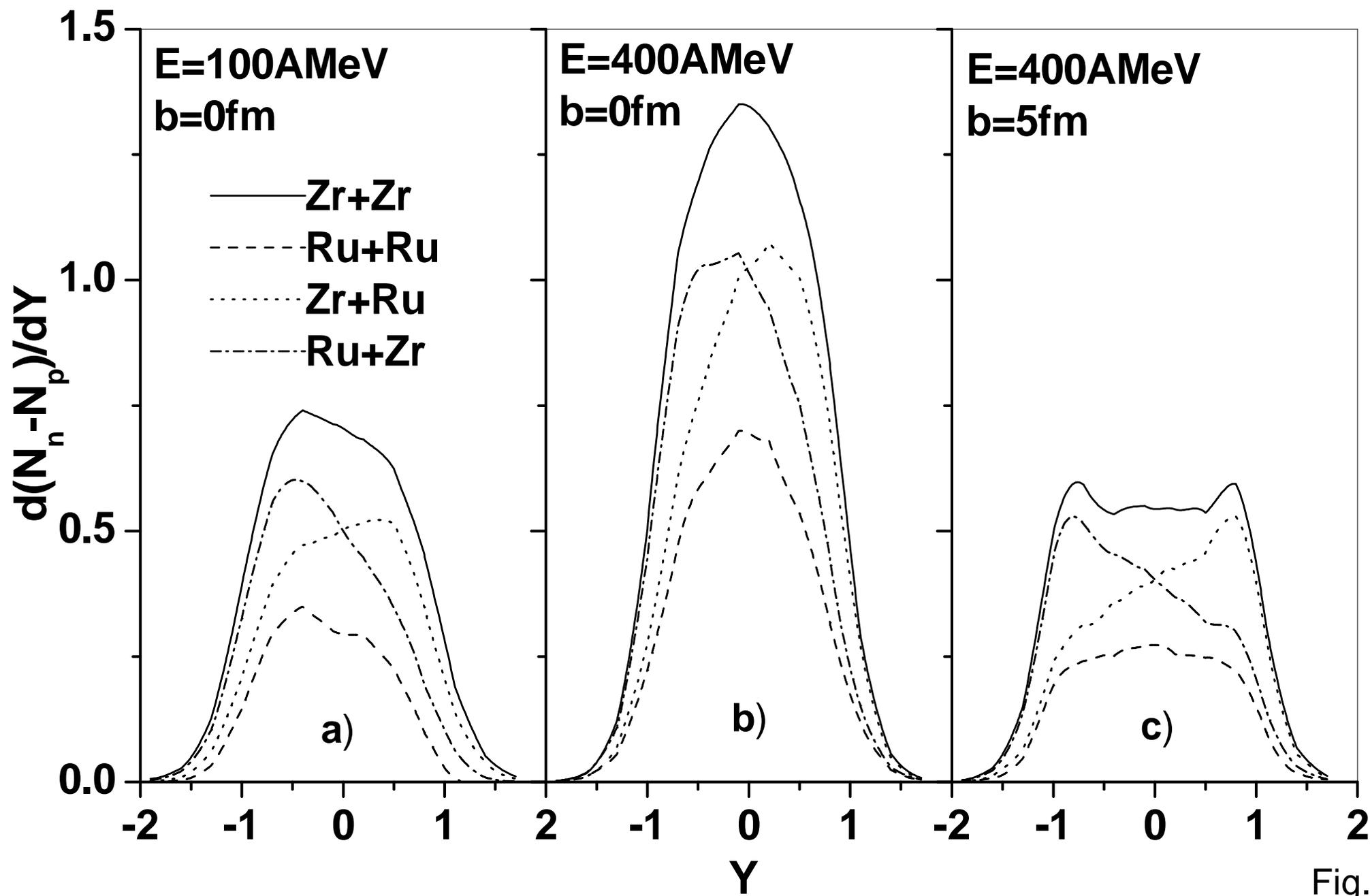
\begin{figure}
\caption{ The neutron-proton differential rapidity distribution for
               the same reactions as Fig. 1
                a) at 100 AMeV, $b=0fm$ and b) at 400 AMeV, $b=0fm$ and c) at 400 AMeV, $b=5fm$.
        }
  
\label{fig3}
\end{figure}

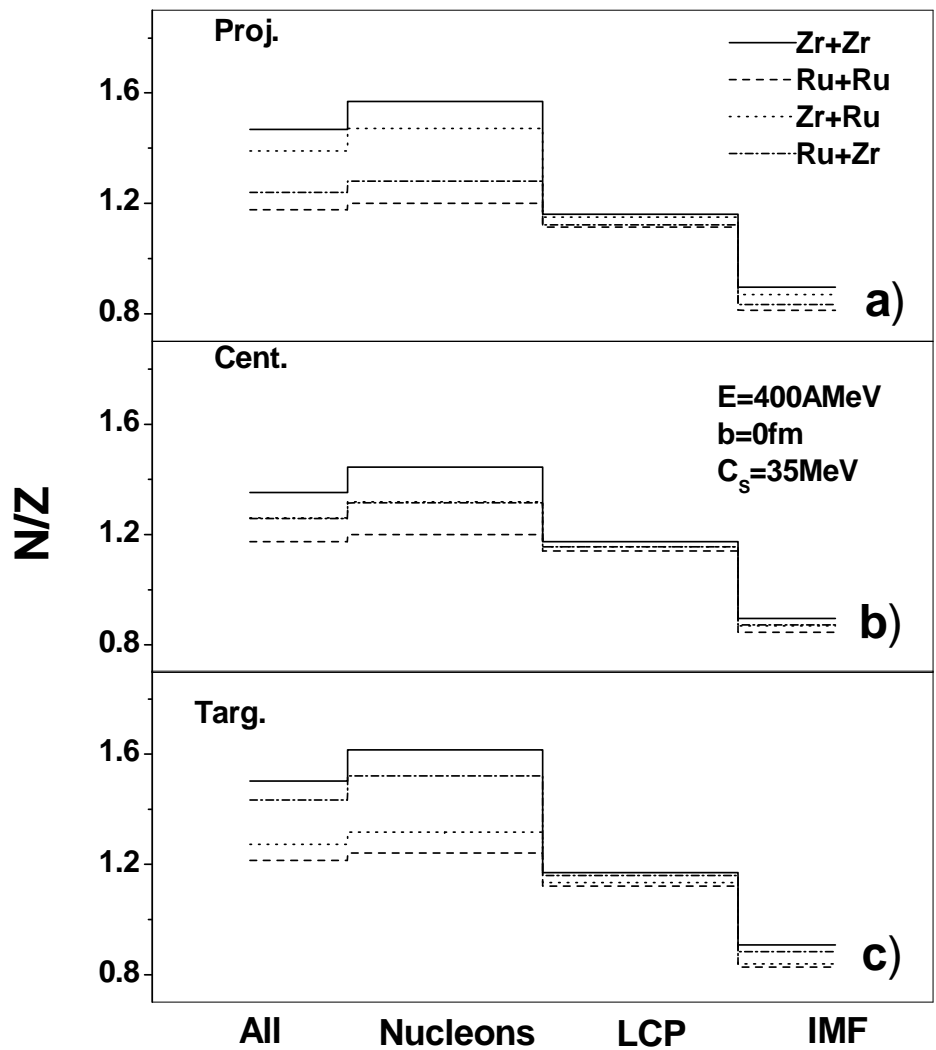
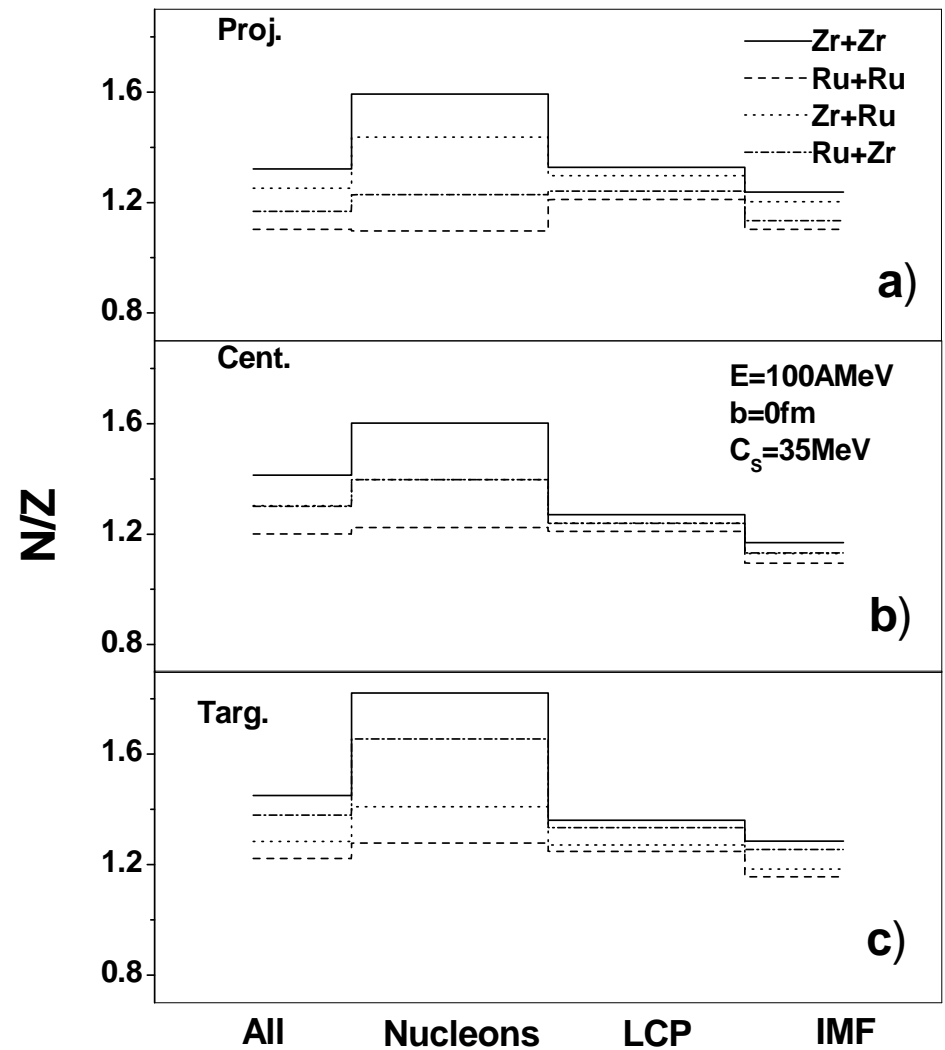
\begin{figure}
\caption{ (I) The average $N/Z$ ratio of emitted nucleons, light 
            charged particles and intermediate mass fragments
                at a) projectile rapidity region,  b) central rapidity region, 
                 c) target rapidity region
            for the same reactions as Fig. 1 at $E=400AMeV$, $b=0fm$. (II)
            The same as (I) but at $E=100AMeV$.
         }
\label{fig4}
\end{figure}

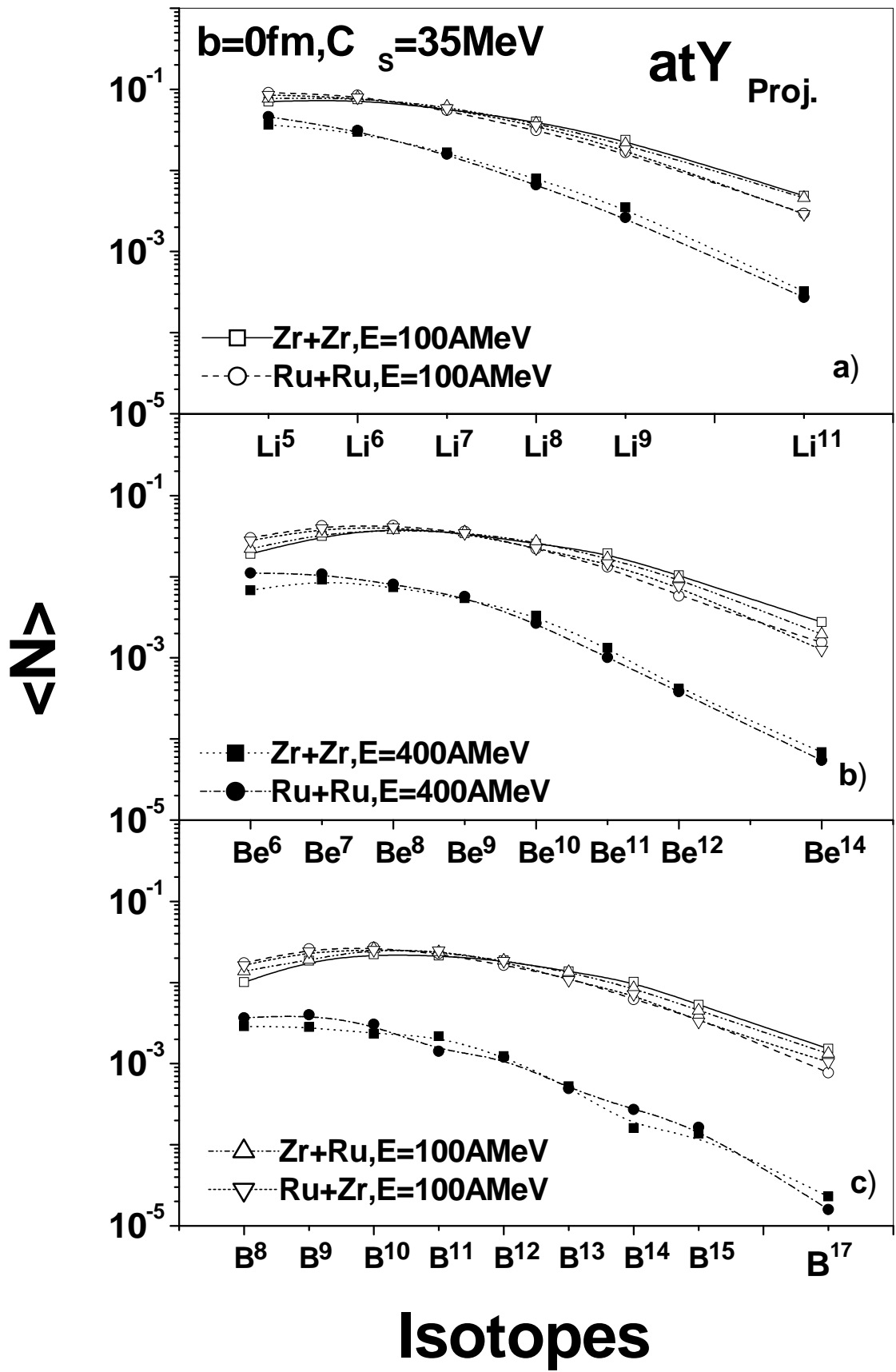
\begin{figure}
\caption{ The isotope distribution of Li, Be, B at projectile region 
            for the same reactions as Fig. 1 at $E=100AMeV$ and $400AMeV$, $b=0fm$, respectively.
        }
\label{fig5}
\end{figure}

\end{document}